\documentclass[lnbip]{svmultln}

\usepackage[small,it]{caption}
\usepackage[small,compact]{titlesec}

\usepackage{graphicx}
\usepackage[ruled,vlined]{algorithm2e}
\usepackage{booktabs}
\usepackage{url}
\usepackage{footnote}

\usepackage{paralist}

\usepackage{graphicx}
\usepackage{subfig}

\usepackage{ifthen}
\usepackage[normalem]{ulem} %
\usepackage{xcolor}

\newboolean{showedits}
\setboolean{showedits}{true} %
\ifthenelse{\boolean{showedits}}
{
	\newcommand{\del}[1]{\textcolor{red}{\sout{#1}}} %
}{
\newcommand{\del}[1]{} %

}
\usepackage{amssymb}
\newboolean{showcomments}
\setboolean{showcomments}{true}
\newcommand{\id}[1]{}

\ifthenelse{\boolean{showcomments}}
{\newcommand{\nb}[2]{
		{\colorbox{orange}{\bfseries\sffamily\scriptsize\textcolor{white}{#1}}}
		{\textcolor{orange}{\sf\small$\blacktriangleright$\textit{#2}$\blacktriangleleft$}}}
	}
{\newcommand{\nb}[2]{}
	\renewcommand{\del}[1]{} %
	}

\usepackage{amsmath}
\usepackage{tikz}

\usepackage{array}
\makeatletter
\newcommand{\thickhline}{%
	\noalign {\ifnum 0=`}\fi \hrule height 1.5pt
	\futurelet \reserved@a \@xhline
}
\makeatother

\usepackage{listings}
\begin{document}

	\mainmatter              %

	\title{Considering Polymorphism in Change-Based Test Suite Reduction}
	\titlerunning{Change-Based Test Suite Reduction}

	\author{Ali Parsai \and Quinten David Soetens \and Alessandro Murgia \and Serge Demeyer}
	\authorrunning{Ali Parsai et al.}   %
	\tocauthor{Ali Parsai,Quinten Soetens,Alessandro Murgia,Serge Demeyer}
	
	\institute{University of Antwerp, Antwerpen, Belgium\\
ali.parsai@student.uantwerpen.be 
\{quinten.soetens;alessandro.murgia;serge.demeyer\}@uantwerpen.be}
	
	\maketitle

\begin{abstract} 
	
	With the increasing popularity of continuous integration, algorithms for selecting the minimal test-suite to cover a given set of changes are in order. 
	This paper reports on how polymorphism can handle false negatives in a previous algorithm which uses method-level changes in the base-code to deduce which tests need to be rerun. We compare the approach with and without polymorphism on two distinct cases ---PMD and CruiseControl--- and discovered an interesting trade-off: incorporating polymorphism results in more relevant tests to be included in the test suite (hence improves accuracy), however comes at the cost of a larger test suite (hence increases the time to run the minimal test-suite). 
	
	\keywords{test selection, unit-testing, change-based test selection, polymorphism, ChEOPSJ}
	
\end{abstract}

\section{Introduction}

The advent of agile processes with their emphasis on test-driven development~\cite{Beck:2002:TDD:579193} and continuous integration~\cite{fowler05integration} implies that developers want (and need) to test their newly changed or modified classes or components early and often~\cite{McGregor2007}. Yet, as Runeson observed in a series of workshops with testing teams, some unit test suites take hours to run~\cite{Runeson2006}. In such a situation, a ``\emph{retest all}'' approach which maximizes the chances of 
verifying if (i) the new functionalities introduced are working properly and (ii) the refactoring of the previous ones do not break the code,
takes too long to provide rapid feedback in a normal edit-compile-run cycle.

A series of interviews we conducted with developers working in different agile teams confirmed that rapid feedback in the presence of a large suite of unit-tests is critically important. When developers address a change-request, they make a chain of changes in the code base, fire a manually selected subset of the unit tests to confirm the system still functions as expected, commit their changes to the code base, run the continuos integration build ---the developers we interviewed reported that a ``retest-all'' takes between 8 and 10 hours--- and in the meantime proceed with the next change request. Most of the time this works fine, but in some occasions the continuous integration build reveals a regression fault and then developers must switch contexts to resolve the fault. One team leader determined that it takes at least 10 minutes before a developer mentally reconstituted the context; since each failed integration build involves several context switches it follows that they easily add an extra half hour just to get a developer in the right frame of mind. Another team leader pointed out that as a system grows and becomes more complex, it is more difficult to identify a suitable test subset hence failed integration builds occur more frequently. A back-of-the-envelope estimation based on their latest quarterly release, revealed that failed integration builds add at least two extra hours per working day.

Essentially, there are three possible strategies to achieve a rapid feedback cycle in the presence of a large suite of unit tests: (a) \emph{parallelisation}, i.e. perform a ``retest all'' on a battery of dedicated test servers to reduce the time to execute the test; (b) \emph{smoke tests}, i.e. define a few representative tests as a quick and dirty verification; (c) \emph{test selection}, i.e. select the subset of the complete test suite covering the last changes made. In this paper, we focus on the latter, however point out that from a pragmatic point of view, a combination of the three strategies is desirable.

Test selection is the problem to ``\emph{determine which test-cases need to be re-executed [\ldots] in order to verify the behavior of modified software}''~\cite{Engstrom:2010:SRR:1645441.1645567}. It has been the subject of intense research in the area of regression testing, however is recently also studied in the context of agile approaches. We refer the interested reader to a survey by Engstr\"{o}m et. al, for an overview of the former~\cite{Engstrom:2010:SRR:1645441.1645567} while Hurdugaci et al.~\cite{Hurdugaci:2012:ASD:2191744.2192546}, Zaidman et al.~\cite{Zaidman2011} are some examples of the latter.%

We ourselves experimented with one particular test selection technique and reported about it during the CSMR 2013 conference~\cite{Soetens2013}. In essence, the algorithm builds a series of dependencies between methods that have been changed ---all of which are captured by the ChEOPSJ tool~\cite{Soetens2012}--- and from that deduces all tests which directly or indirectly invoke those methods. Our results showed that given a list of methods which changed since the latest commit, we could select a subset of the entire test suite which is significantly smaller. The selected subset is not safe as it occasionally misses a few relevant tests. However it is \emph{adequate} since the test-coverage ---expressed as ``percentage of mutations that were killed''~\cite{Andrews2005})--- remained the same.

Nevertheless, the algorithm explained in~\cite{Soetens2013} made one simplifying assumption, namely that developers would refrain from using polymorphism, i.e. invocations of overridden methods, abstract methods or methods declared in interfaces~\cite[Ch. 2]{Booch2006}. This simplifying assumption did not hold in one of the cases (namely PMD) and as a result our algorithm missed several relevant tests.
For this reason, we decided to repeat the previous experiment to address the following research question:

\vspace{\baselineskip}
\textbf{RQ.} \textit{Does considering polymorphism improve the quality of the reduced test suite in a realistic situation?}
\vspace{\baselineskip}

In this experiment, we applied the improved algorithm on the two cases used in the original experiment: PMD and CruiseControl. %

The rest of this paper is structured as follows. In section~\ref{newsection2}, we describe the approach for test selection and the test selection algorithm. In section~\ref{EXPERIMENT} we explain the experimental setup. In section~\ref{ANALYSIS}, we present the results of the experiment. 
In section \ref{ThreatsToValidity}, we describe which factors may jeopardize the validity of our analysis. In section \ref{RelatedWork}, we summarize the related work. Finally, in section~\ref{CONCLUSION} we wrap up the work with a summary and conclusions. %

\section{Supporting code change and test selection}
\label{newsection2}
This section describes how we introduced the concept of polymorphism in the test suite reduction algorithm of ChEOPSJ. We start by introducing ChEOPSJ and then its test selection algorithm\footnote{ The proposed procedure is easily generalizable for any object oriented 
	system. There are commercial tools (e.g. Visual Studio) which try to solve the same problem for a specific language, but does not publicly provide the used technique.}.

\begin{figure}
	\centering
	\includegraphics[width=0.62\textwidth]{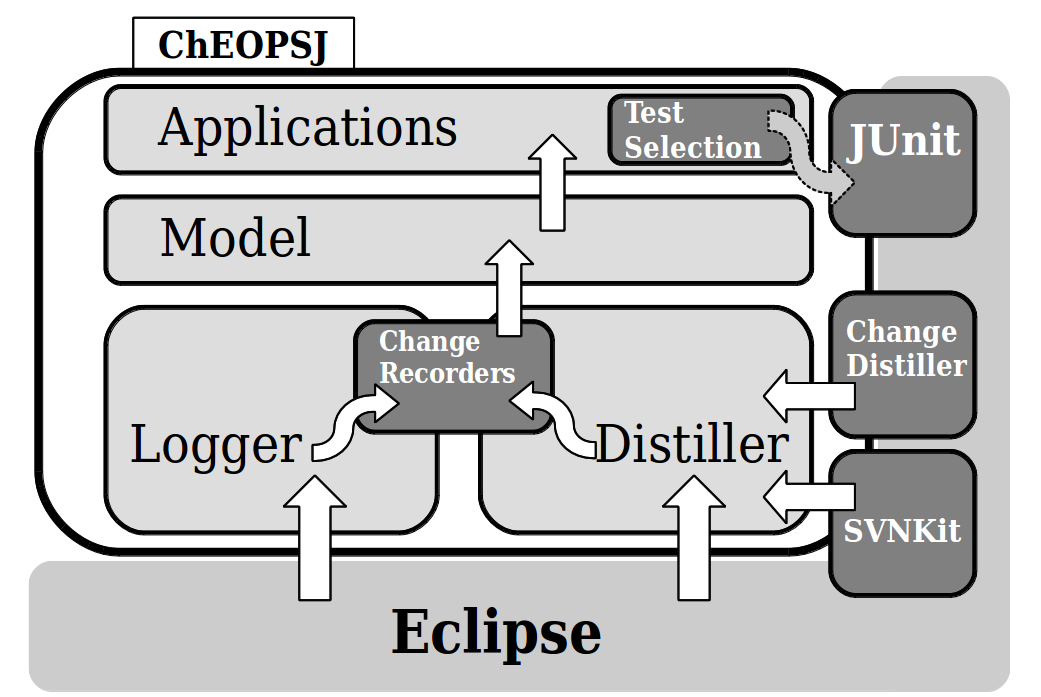}
	\caption{The layered design of ChEOPSJ}
	\label{cheopsjinternals}
\end{figure}

ChEOPSJ\footnote{The acronym ChEOPSJ stands for: Change ad Evolution Oriented Programming Support for Java.} is a proof of concept system able to extract and model software changes.
This tool is implemented as a series of Eclipse plugins. Figure~\ref{cheopsjinternals} shows the overview of its structure.
At the center of the tool we have a plugin that contains and maintains the change model. There are two plugins that are responsible for populating the change model. The \emph{Logger} generates change objects by recording actions a developer makes in the main editor during a development session, while the \emph{Distiller} obtains the change objects by mining a Subversion repository.  Once the change model is populated with first class change objects, many applications can be built on top of ChEOPSJ that can use them for their own purpose. Our TestSelection plugin is one such application.

Our approach for change-based test selection uses the first class change-objects of ChEOPSJ. We define a \emph{Change} as an object representing an action that changes a software system. As such, a change becomes a tangible entity that we can analyze and manipulate. We define three kinds of \textit{Atomic Changes}: \texttt{Add}, \texttt{Modify} and \texttt{Remove}. These changes act upon a \textit{Subject} representing an actual source code entity. For these subjects we can use any model that is capable of representing source code entities. We chose the FAMIX model as defined in~\cite{Demeyer1999} since its model is usable to describe many object oriented programming languages. As such our approach is applicable in any object oriented setting. 

Our model also defines dependencies between the changes. These are deduced from the FAMIX model, which imposes a number of invariants to which each model must adhere. For instance, there is an invariant that states that each method needs to be contained in a class. This means that there is a precondition for the change ($m_{Add}$) that adds a method $m$ to a class $c$. There should exist a change ($c_{Add}$) that adds the class $c$ and there is no change that removes the class $c$. We can then say that the change $m_{Add}$ depends on the change $c_{Add}$. More generally we can say that a change object $c_1$ depends on another change object $c_2$ if the application of $c_1$ without $c_2$ would violate the system invariants. 

The change based test selection heavily relies on these dependencies, as it traces them from a selected change to the additions of test methods. To calculate the reduced test suites we execute Algorithm~\ref{alg:testselect}.
\begin{algorithm}
	\scriptsize
	\caption{select relevant tests}
	\label{alg:testselect}
	
	\KwIn{A $ChangeModel$, A set $SelectedChanges$}
	\KwOut{A Map that maps each selected change to a set of relevant tests.}
	
	\ForEach{$c$ in $SelectedChanges$}{
		$calledMethod$ = findMethodAddition(hierarchicalDependencies($c$)); \\
		$invocations$ = invocationalDependees($calledMethod$);\\
		\ForEach{$i$ in $invocations$}{
			invokedBy = findMethodAddition(hierarchicalDependencies($i$)); \\
			\ForEach{$m$ in $invokedBy$}{
				\If{ $m$ is a test }{
					add $m$ to $relevantTests$;
				}
				\Else{
					\If{$m$ was not previously analyzed}{
						$tests$ = selectRelevantTests($m$);\\
						add $tests$ to $relevantTests$;
					}
				}
			}
		}
		map $c$ to $relevantTests$;
	}
\end{algorithm}

In this algorithm, we iterate all selected changes and map each change to their set of relevant tests. 
We start by finding the change that adds the method in which the change was performed. We can find this change, by following the chain of hierarchical dependencies and stop at the change that adds a method. In Algorithm~\ref{alg:testselect} this is presented by a call to the procedure \texttt{findMethodAddition}. After this call \texttt{calledMethod} will be the change that adds the method in which the change \texttt{c} took place. 
Next we need to find all changes that add an invocation to this method. These are found by looking for \texttt{invocationalDependencies}. For each of these changes, we again look for the change that adds the method in which these invocations were added. And thus we find the set of all changes that add a method that invokes the method that contains our selected change. We then iterate these method additions and check whether these changes added a test method. If this was the case we consider this test method as a relevant test for the originally selected change. If on the other hand the added method was not a test method, then we need to find the relevant tests of this method and that set of tests needs to be added to the set of relevant tests for the selected change. 

\paragraph{Polymorphism during test selection.} 
In our original approach, the change model assumed that invocations were a one to one relationship between the caller and the callee. As such the addition of an invocation was dependant on the addition of the caller method as well as on the addition of the callee method. We could statically determine the latter based on the type of the variable on which the method was invoked. However with polymorphism this is not necessarily the case, as a method invocation might invoke any of a number of possible methods. 

\begin{figure}
	\centering
	\begin{minipage}{.34\textwidth}
		\begin{lstlisting}
		class Foo{
		public void foo(){
		}
		}
		
		class Bar extends Foo{
		public void foo(){
		}
		}
		
		class FooBarTest{
		private Foo f;
		public void SetUp(){
		f = new Bar();
		}
		public void fooTest(){
		f.foo();
		}
		}\end{lstlisting}
	\end{minipage}
	\hspace{1mm}
	\begin{minipage}{.62\textwidth}
		\includegraphics[width=\textwidth]{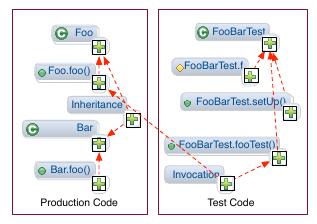}
	\end{minipage}
	\caption{Example of changes with polymorphic call.}
	\label{polymorphic}
\end{figure}

Take for instance the code in Figure~\ref{polymorphic}, here we have a class \texttt{Foo} that declares a method \texttt{foo} and a subclass \texttt{Bar} that declares a polymorphic version of that same method. Our test invokes the method \texttt{foo} on a variable \texttt{f} of type \texttt{Foo}, hence our algorithm would state that this test is relevant for all changes in the method \texttt{Foo.foo()}. However in the \texttt{setUp} method, the variable \texttt{f} is instantiated as an object of type \texttt{Bar} so this test is in fact also relevant for the method \texttt{Bar.foo()}, which is a link that our test selection algorithm missed. Hence our algorithm did not take into account actual methods that are invoked at runtime like polymorphic methods, abstract methods or methods declared in interfaces. 

\begin{figure}
	\centering
	\begin{minipage}{.64\textwidth}
		\includegraphics[width=\textwidth]{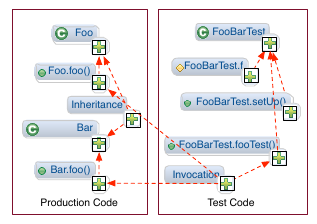}
	\end{minipage}
	\caption{Updated model with dependency from invocation to all possible methods with same identifer.}
	\label{polymorphicupdate}
\end{figure}

As a simple workaround, we slightly changed our change model so that an addition of an invocation is dependant of all additions of methods that this invocation can possibly be referring to, based on its identifier and parameter list. So when a method invocation is added, this addition now depends on all method additions that add a method with this same identifier. This would change the model of the changes in Figure~\ref{polymorphic} to the model represented in Figure~\ref{polymorphicupdate}. Note that in the new model, there is an added dependency from addition of the invocation in the test method to the addition of the method \texttt{Bar.foo}. So now our test selection algorithm will say that the test \texttt{FooBarTest.fooTest()} is relevant for changes in both the methods \texttt{Foo.foo()} and \texttt{Bar.foo()}.

\begin{figure}
	\centering
	\begin{minipage}[b]{0.9\textwidth}
		\centering
		\includegraphics[width=\textwidth]{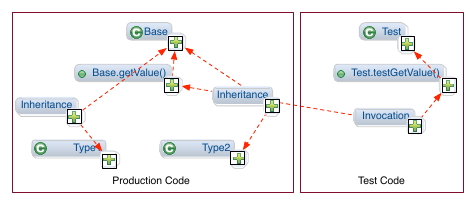}
		\nocaption{Before refactoring}
	\end{minipage}
	
	\begin{minipage}[b]{0.9\textwidth}
		\centering
		\includegraphics[width=\textwidth]{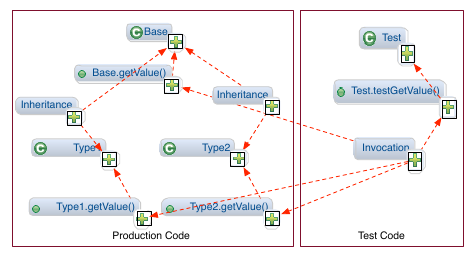}
		\nocaption{After refactoring}
	\end{minipage}

	\caption{Model state before and after the refactoring}
	\label{refactoringexample-model}
\end{figure}

We report another example in Figure~\ref{refactoringexample}: the refactoring \textsc{ReplaceConditionalWithPolymorphism}. In this example the class \texttt{Base} contains a method \texttt{getValue} which uses a conditional to determine its actual runtime type and based on that will perform a different action. The refactoring then involves the creation of polymorphic versions of this method in the two subtypes that perform the type specific actions. We show the change model of this code before and after the refactoring in Figures~\ref{refactoringexample-model}. The test of this code, invokes the method \texttt{getValue} on the superclass, which results in a dependency from the test to the addition of that method. This means that this test is relevant to all changes in the \texttt{getValue()} method. In the version before refactoring this would be correct, however in the post-refactored class without polymorphic support, a test containing an invocation of \emph{getValue} on an object of class \emph{Base} will not be selected for objects of types \emph{Type1} and \emph{Type2}; Because, the addition of the invocation of \emph{getValue} in class \emph{Test} is dependent on the addition of the abstract method \emph{Base.getValue} and not the addition of methods \emph{Type1.getValue} and \emph{Type2.getValue}. Whereas, with the polymorphic support \emph{Test} will be selected, because as shown in Figure~\ref{refactoringexample-model} there is a dependency from the invocation to all methods with the identifier ``getValue''.

\begin{figure}
	\centering
	\begin{minipage}{.40\textwidth}
		\begin{lstlisting}
		class Base{
		int getValue(){
		switch (_type){
		case Type1:
		return getType1Value();
		case Type2:
		return getType2Value();
		}
		}
		}
		
		class Type1 extends Base{
		}
		
		class Type2 extends Base{
		}
		\end{lstlisting}
	\end{minipage}
	\hspace{0.5cm}
	\begin{minipage}{.40\textwidth}
		\begin{lstlisting}
		class Base{
		int getValue();
		}
		
		class Type1 extends Base{
		int getValue(){
		return getType1Value();
		}
		}
		
		class Type2 extends Base{
		int getValue(){
		return getType2Value();
		}
		}
		\end{lstlisting}
	\end{minipage}
	\caption{Replace Conditional with Polymorphism refactoring}
	\label{refactoringexample}
\end{figure}

\section{Experimental Setup}
\label{EXPERIMENT}
We use mutation testing to estimate the real-life behavior of the reduced test suite. 
We replicated a previous study to evaluate the benefits of the introduction of polymorphism~\cite{Soetens2013} . 

\subsection{Mutation Testing}

Mutation testing\footnote{The interested reader may refer to~\cite[Ch. 7]{Dasso2007} for more information regarding mutation testing.} provides a workaround to measure the quality of a test suite and identify its weak points~\cite{Andrews2005}. 
In mutation testing  intentional faults are put inside a %
fault-free program by applying mutation operators (or mutators). Mutators are chosen according to a fault model
so that the generated faults correspond to a realistic situation. 
A mutation is \textit{killed} if it causes a test to fail, while if it does not fail any tests, it has \textit{survived} the experiment. We can then consider the mutation coverage, which is a ratio between the number of mutants that were killed over the number of mutants that were introduced. Mutation coverage provides a reliable metric to measure the quality of a test suite~\cite{Andrews2005}. A higher mutation coverage, means that more of the introduced mutants were killed and consequently that your test suite is of better quality.

In the previous experiment, PIT\footnote{\url{http://pitest.org}} was used as the main means for mutation testing. 
In this experiment, we use the same configuration of PIT used in the previous study~\cite{Soetens2013}.
PIT provides byte code mutation testing by integration into the build procedure ---either Ant or Maven---
of the target software. 
To get a base measurement of the quality of the test suites, %
PIT is run considering all classes and the full test suite.
Then separate build files were generated for each class which included only the tests ChEOPSJ deemed relevant to the class in question. 
We compare the mutation coverage of each reduced test suite with the mutation coverage of the full test suite by looking  at the mutants that survived the reduced suite but that were killed in the full suite. Ideally the mutation coverage of the reduced test suite should equal the mutation coverage of the entire test suite. When the mutation coverage is lower, it means that we have missed some relevant tests in our selection.

The results of this mutation coverage analysis (using %
polymorphism) %
is then compared to the results of our previous experiment (where polymorphism was not taken into account).

\subsection{Selected Cases}

To be able to measure the impact of supporting polymorphism in ChEOPSJ, we examined the same cases (PMD\footnote{\url{http://pmd.sourceforge.net}} and CruiseControl\footnote{\url{http://cruisecontrol.sourceforge.net}}) as the previous study~\cite{Soetens2013}. 
Moreover, we use the same revisions of both projects %
to reliably repeat the experiment.
CruiseControl %
is a continuous integration tool and an extensible framework for creating a custom continuous build process. %
PMD %
is a source code analyzer which finds common programming flaws like unused variables, empty catch blocks, unnecessary object creation. %
These projects are open-source, written in Java and accessible through SVN. %

The sizes of these projects and the selected revisions are shown in Table~\ref{tab:sizes}. 

\begin{table}[htbp]
	\centering
	\begin{smaller}
		\begin{tabular}{c|cccccc}
			\hline
			\toprule
			Project & Version & Src & Src & Test & Test & Build \\ 
			& analyzed & KLOC & NOC & KLOC & NOC & Process\\ 
			\midrule
			Cruisecontrol & rev. 4601 & 26.5 & 376   & 24.5 & 295 & ant \\
			PMD                & rev. 7706 & 46     & 804   & 9       & 215 & maven\\  \bottomrule
		\end{tabular}
	\end{smaller}
	\caption{Number of 1000 Lines of Code (KLOC) and Number of Classes (NOC) for both source code and test code (measured with InFusion 7.2.7).}
	\label{tab:sizes}
\end{table}

\section{Results and Discussion}
\label{ANALYSIS}
This section analyses the results of the test selection algorithm. We compute for all classes ---in the full test suite 
and in the reduced one---  the mutants generated. Then, we count the number of mutants killed and the number of classes involved. Moreover, we compare our results with those achieved with the previous version of ChEOPSJ.

\subsection{PMD}
The test suite of PMD covers 665 classes and with PIT we generate mutants on each one of them. Using PIT on the reduced test suite, we generated mutants for 607 classes\footnote{PIT does not generate mutations in a class if the given test suite has no coverage over that class.}. When comparing this to the previous experiment we have a significant improvement, as the previous version generated mutants for only 144 classes. 

\begin{figure}
	\centering
	\subfloat[][Percentage of classes with\\improved mutation coverage]{ %
		\includegraphics[width=0.49\textwidth]{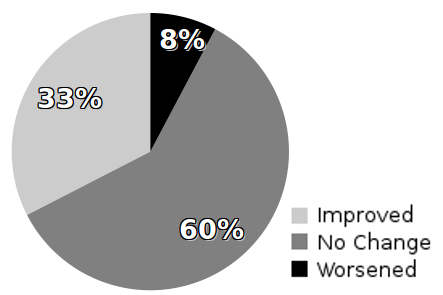}
		\label{pmddifferenceinmutationcoverage}
	}%
	\subfloat[][Mutation coverage comparison\\between all tests and selected tests]{%
		\includegraphics[width=0.49\textwidth]{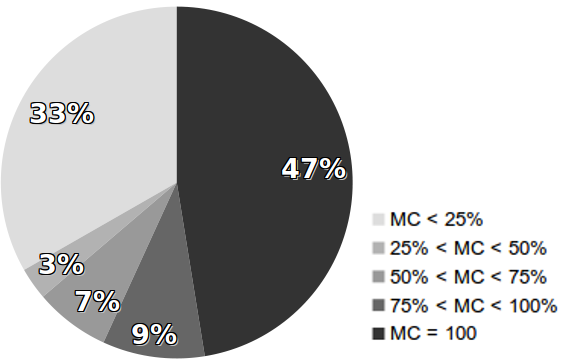}
		\label{pmdmutationcoverage}
	}%

	\caption{Mutation coverage on PMD}
	
\end{figure}

In  Figure~\ref{pmdmutationcoverage} we compare the mutation coverage of the reduced test suites with the full test suite. We find that the reduced test suites of 47\% of the classes have the same mutation coverage as the full test suite. This matches the results of the previous experiment~\cite{Soetens2013}, where there were 50\% of reduced test suites that had an equal number of mutants killed compared to the mutation coverage of the full test suite.

We also observed another improvement with respect to the experiment made with the previous version of ChEOPSJ. 
In both experiments, the reduced test suites have 128 common classes with 4908 mutations. 
In the previous experiment the reduced test suites killed 2114 mutants, whereas in our current experiment the reduced test suites killed a total of 2327 mutants. This means that our improved approach killed 213 more mutants than before. So considering polymorphism resulted in an improvement of 4.3\% in the quality of the reduced test suite. In Figure~\ref{pmddifferenceinmutationcoverage} we report the percentage of classes with improved mutation coverage. 
In the case of PMD, 33\% of the classes have improved mutation coverage while 8\% have worsened coverage. 

\begin{figure}[htbp]
	\centering
	\includegraphics[width=\textwidth]{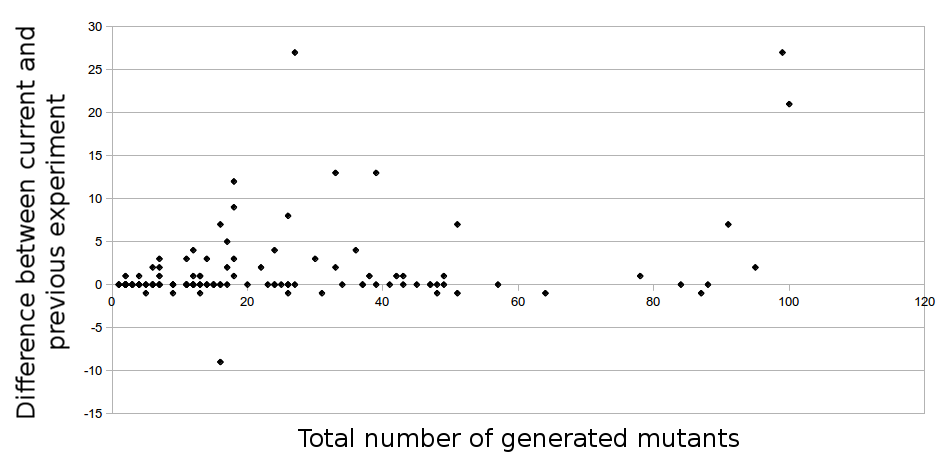}
	\caption{Difference in killed mutants for PMD}
	\label{pmdsurvivedmutants}
\end{figure}

Finally, to inquire to what extent there is an improvement on the number of killed mutants, we compare them with respect to the total number of mutants generated for any class. This is reported in Figure ~\ref{pmdsurvivedmutants} where the X axis is the number of generated mutants and the Y axis is the difference in number of killed mutants for each class in the two experiments. Given a class, we have an improvement of the test suite reduction whenever the number of killed mutants is higher than before. Therefore, the accumulation of points near the Y axis means that those classes with small number of mutants have been impacted more and their mutation coverage is significantly improved. %

\subsection{CruiseControl}

\begin{figure}
	\centering
	\subfloat[][Percentage of classes with\\improved mutation coverage]{ %
		\includegraphics[width=0.49\textwidth]{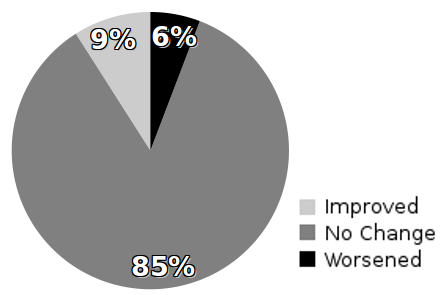}
		\label{cruisecontroldifferenceinmutationcoverage}
	}%
	\subfloat[][Mutation coverage comparison\\between all tests and selected tests]{%
		\includegraphics[width=0.49\textwidth]{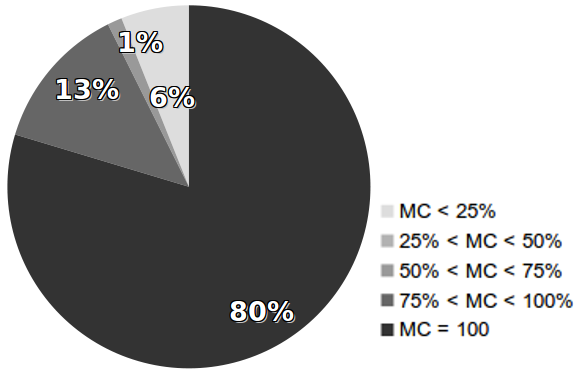}
		\label{cruisecontrolmutationcoverage}
	}%

	\caption{Mutation coverage on CruiseControl}
	
\end{figure}

PIT generates mutants in 246 classes covered by the full test suite of CruiseControl and in 231 classes covered by the reduced test suites. From a total of 6860 mutants generated,  3627 were killed. This is the same number of generated mutants as in the previous experiment and the new reduced test suites kill only 4 more mutants than the old reduced test suites. This means that the addition of polymorphism had nearly no effect on the results in the case of CruiseControl.
This is also confirmed by  Figure~\ref{cruisecontroldifferenceinmutationcoverage}. 
This Figure shows that the percentage of classes with improved mutation coverage is 9\% compared to 6\% for classes with worsened mutation coverage. Also, almost all of the classes have the same mutation coverage as before; and only one class has a significantly better coverage than before. On the other hand, the quality of the whole test suite remains similar to the previous experiment when 80\% of classes have the exact same mutation coverage as running the whole test suite as can be seen in Figure~\ref{cruisecontrolmutationcoverage}.  

\begin{figure}[htbp]
	\centering
	\includegraphics[width=0.9\textwidth]{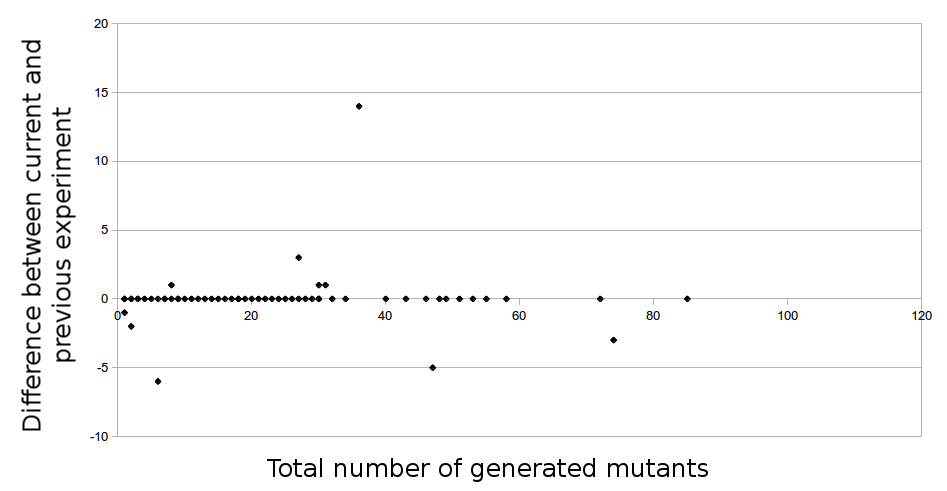}
	\caption{Difference in killed mutants for CruiseControl}
	\label{cruisecontrolsurvivedmutants}
\end{figure}

Figure \ref{cruisecontrolsurvivedmutants} reports the difference in number of killed mutants in the two experiments for each class by the total number of mutants generated. As we can see the quality of the test suites remain the same with few exceptions. Meanwhile, those classes with worsened coverage have only slightly less mutation coverage than in the previous experiment.

\textbf{RQ.} \textit{Does considering polymorphism improve the quality of the reduced test suite in any realistic situation?}\\
In a realistic situation the effect of the addition of polymorphism to the test selection must be considered case-by-case.
In the case of PMD, the project's heavy reliance on polymorphic structures means that the results for the mutation coverage have improved greatly by considering this concept. However, there are a lot of tests that are not being selected for different reasons. For example, PMD uses XML files as input for \textit{rule-based tests}. %
This kind of tests %
do not result in any invocations, and therefore are not detected and selected by ChEOPSJ. As a consequence, the results are less than optimal, considering the fact that such tests are a huge part of the whole test suite. 
In the case of 
CruiseControl, the effects on the mutation coverage are minimal. This is due to two reasons: (i) the project does not use polymorphism extensively, and %
(ii) in the original experiment the mutation coverage was already  good \cite{Soetens2013}. 

\begin{figure}
	\centering
	\begin{minipage}[b]{0.45\textwidth}
		\centering
		\includegraphics[width=\textwidth]{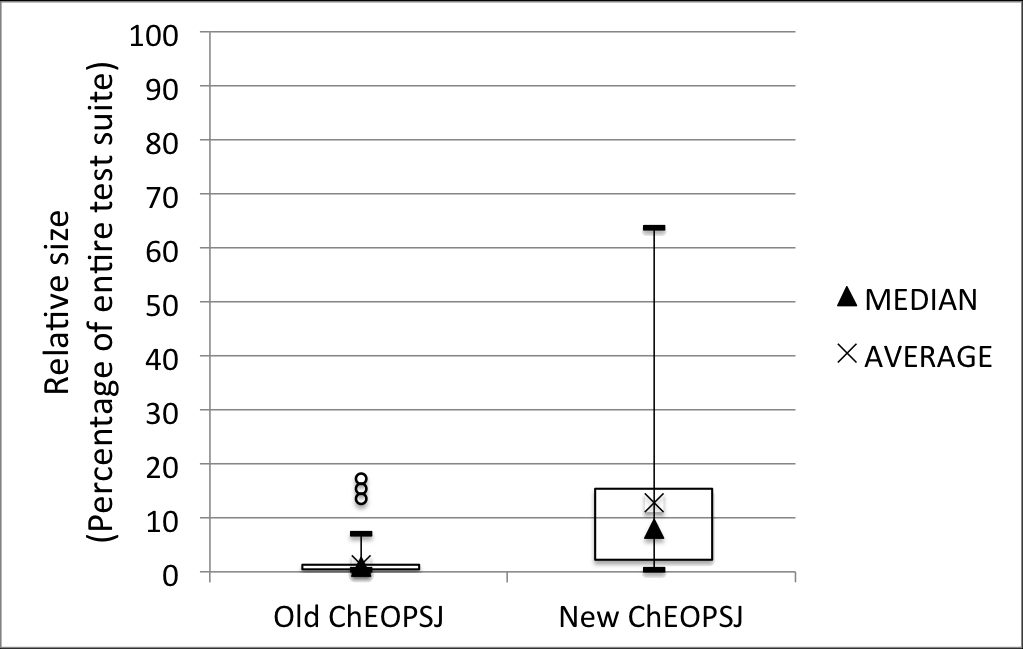}
		\nocaption{PMD}
	\end{minipage}%
	\hspace{2mm}
	\begin{minipage}[b]{0.45\textwidth}
		\centering
		
		\includegraphics[width=\textwidth]{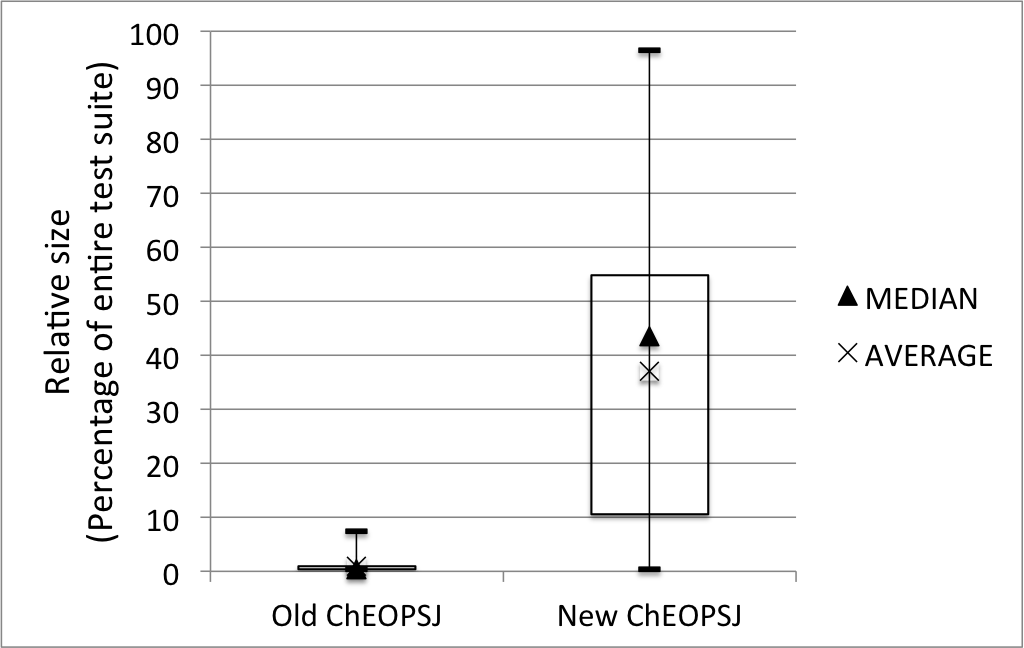}
		\nocaption{CruiseControl}
	\end{minipage}%
	\caption{Comparison of reduced test suite size between previous and current experiments for CruiseControl and PMD.}
	
	\label{sizecomparison}	 
\end{figure}

For PMD and CruiseControl, we compute  the test size reduction as the percentage of test classes in the selected subset against the number of test classes in the entire test suite (Figure~\ref{sizecomparison}). In both cases, we observe that the size of the reduced test suite is much larger
when the polymorphism is considered. %
This means that there is a trade-off between size reduction and considering polymorphism. The size of this trade-off is determined on a case-by-case basis. For example, in the case of PMD, the reduced test suite is still good enough to be useful in a realistic situation. However, in CruiseControl the reduction in size may not be enough to promote the adoption of polymorphism. 

As conclusion, to improve the quality of test selection,  the adoption of polymorphism should be provided as an option. %
However, if the developer does not have enough knowledge of software characteristics, a workaround would be the creation of a heuristic function 
that detects the reliance of the software system on polymorphism. %

\section{Threats to Validity}
\label{ThreatsToValidity}

In this section we present the threats to validity of our study according to the guidelines reported in \cite{robert2003case}. 

Threats to internal validity concern confounding factors that can influence the obtained results. In this study we used the code base of ChEOPSJ which does not include  constructor invocations. As a consequence we may erroneously miss relevant tests. We could fix this problem incorporating these language constructs in the change model of ChEOPSJ.  %

Threats to construct validity focus on how accurately the observations describe the phenomena of interest. For our experiment, the elements of interest are (1) the test suite reduction and (2) the number of missing faults due to not-retested code. We measure the first one as the ratio between number test classes  in the reduced test suite versus the complete test suite. The second one is computed as the number of mutants killed. Both approaches are used in literature for the same purpose \cite{Soetens2013}. However, other methods are suitable for evaluating the test suite reduction and number of missing faults.

Threats to external validity correspond to the generalizability of our experimental results. We use the projects CruiseControl and PMD. Even if both systems are sufficiently different, yet more projects are necessary to generalize our findings. 

Threats to reliability validity correspond to the degree to which the result dependent on the used tools. To implement the algorithm we use the baseline offered by ChEOPSJ (which relies on Eclipse's internal Java mode) and ChangeDistiller. Both systems are reliable and used to perform research studies \cite{Fluri:2007:CDT:1314036.1314081,Soetens2013}. For the mutation testing we used PIT. This system is actively being developed and improved and can be considered reliable. %

\section{Related Work}
\label{RelatedWork}

Regression testing aims to test code changes to ensure that the correct behavior of the system is preserved \cite{Rothermel:1996:ART:235681.235682}.
In this context, test suite reduction is crucial in continuous integration environments or test-driven development 
\cite{Beck:2002:TDD:579193,fowler05integration}, namely whenever software development has frequent re-testing activities. Regression testing is also useful during code refactoring since refactoring may reverberate on the test suite \cite{Hayes:2009:TTT:1556908.1557013,vanDeursenMoonen2002}. 

Test suite reduction is an active research field \cite{Engstrom:2010:SRR:1645441.1645567,Engstrom:2008:EER:1414004.1414011}. 
It mainly focuses on evaluating the trade-offs between executing a subset of tests and the risk of missing some faults
sneaked into not-retested code sections.
The test selection problem has been handled in many different manners. In the context of static approaches, common used heuristics are naming conventions, fixture element types,  static call graph, lexical analysis, co-evolution \cite{Rompaey:2009:ETL:1545011.1545440}. Nevertheless, none of previous approaches 
handle shortcomings related to code polymorphism.
In this work, we explore this aspect evaluating how test suite reduction algorithms may be adapted to deal with invocations of polymorphic methods, abstract methods and methods declared in interfaces. 

Integration of the testing activity within the IDE environment is critically important  to achieve a ``continuous testing" system \cite{SaffE2004:ISSTA}. Moreover, providing ad-hoc plugins for test selection is already common in academic research (e.g. TestNForce \cite{Hurdugaci:2012:ASD:2191744.2192546}) or among commercial vendors (e.g. Visual Studio's Test Impact Analysis). 
The second author  %
embedded a test selection algorithms within the Eclipse's plugin and performed few empirical studies to show its performances \cite{Soetens2012, Soetens2013}.
In this work, we extended this plugin to make it able to deal with polymorphism. 

\section{Conclusion}
\label{CONCLUSION}

We replicated an experiment that we did in~\cite{Soetens2013} to analyzes the effects of polymorphism for test suite reduction using a change-based model.

Our goal was to answer the research question: %

\textbf{RQ.} \textit{Does considering polymorphism improve the quality of the reduced test suite in a realistic situation?}

Our results show that polymorphism may have different relevance on PMD and CruiseControl from the point of view of mutation coverage analysis and test suite reduction.

In PMD, one third of the classes had an improved mutation coverage. Relevant tests were found for more classes than before with the same statistical probability for killing mutants.
Overall, a 4\% increase in the rate of total killed mutants is observed. Having the possibility to retrieve \textit{rule-based tests}, our results would probably be better.

On the other hand, for CruiseControl the differences between the two experiments were minimal. This can be attributed to the fact that PMD uses polymorphism in the code extensively and there are some abstract core entities which are used throughout the whole project. The use of polymorphism in CruiseControl is limited and therefore the effects of considering polymorphism would remain minimal. 

From the point of view of the test suite reduction the introduction of polymorphism increases the size of the test suite.
This is a normal trade-off we have to accept if we increase the number of potential relevant tests of our suite. 
A possible workaround would be to determine the level of adaption of polymorphism in the project and then decide if it is valuable to include it during test suite reduction. As such the adoption of polymorphism in the test selection process should be provided as an optional feature for the developers to choose from. 

\noindent\textbf{Contributions.} %
We made the following contributions:
\begin{itemize}
	\item We improved our tool prototype ChEOPSJ to incorporate polymorphism.
	\item Using this improved platform we replicated a previous experiment and confirmed our previous findings.
	\item Finally, we found that there is a tradeoff between the accuracy of our approach and the size of the reduced test suite.
\end{itemize}

\noindent\textbf{Future Work.} 
We will improve the support of common architectural design concepts that are used widely in software systems. We will focus on test generation techniques that use XML specifications and polymorphic tests which use the same abstract entities to provide many tests.

Additionally we will perform more replications in an industrial setting. Will developers be more inclined to run their developer tests more frequently with test selection enabled? Will this result in fewer (regression) faults later in the life-cycle? This to assess the real significance of test selection in a realistic scenario.

\section*{Acknowledgments} 
\small{This work is sponsored by the Institute for the Promotion of Innovation through Science and Technology in Flanders through a project entitled Change-centric Quality Assurance (CHAQ) with number 120028.

\bibliographystyle{plain}
\bibliography{main}

\end{document}